\documentstyle[prb,aps]{revtex}

\begin{document}

\title{Rate of steady--state reconnection in an incompressible
plasma}
\author{Nikolai V. Erkaev}
\address{Institute of Computational
Modelling, Russian Academy of Sciences, 660036 Krasnoyarsk 36,
Russia} \author{Vladimir S. Semenov, Ilya V. Alexeev}
\address{Institute of Physics,
University of St. Petersburg, St. Petergof, 198504 , Russia }
\author{Helfried K. Biernat}
\address{Space Research Institute, Austrian Academy of Sciences,
Schmiedlstrasse 6, A--8042 Graz, Austria}

\maketitle

\begin{abstract}
The reconnection rate is obtained for the simplest case of 2D
symmetric reconnection  in an incompressible plasma. In the short
note (Erkaev et al.\cite{Erkaev}), the reconnection rate is found
by matching the outer Petschek solution and the inner diffusion
region solution. Here  the details of the numerical simulation of
the diffusion region are presented and  the asymptotic procedure
which is used for deriving the reconnection rate is described. The
reconnection rate is obtained as a decreasing function of the
diffusion region length. For a sufficiently large diffusion region
scale, the reconnection rate becomes close to that obtained in the
Sweet--Parker solution with the inverse square root dependence on
the magnetic Reynolds number $Re_m$, determined for the global
size of the current sheet. On the other hand, for a small
diffusion region length scale, the reconnection rate turns out to
be very similar to that obtained in the Petschek  model with a
logarithmic dependence on the magnetic Reynolds number $Re_m$.
This means that the Petschek regime seems to be possible only in
the case of a strongly localized conductivity corresponding to a
small scale of the diffusion region.
\end{abstract}

\section{Introduction}

Magnetic reconnection is a physical process in plasmas which
changes a magnetic field topology and releases stored magnetic
energy. It is one of the central concerns in astrophysical, solar,
space, fusion and laboratory plasmas (e.g., Hones\cite{Hones};
Priest\cite{Priest}).

A key question arising in the reconnection theory is that of the
reconnection rate. So far there are two different
magnetohydrodynamic (MHD) models of reconnection based on the
Sweet--Parker (pure diffusion) (see Parker\cite{Parker57};
Sweet\cite{Sweet}) and the Petschek (slow shock energy conversion)
(see Petschek\cite{Petschek}) approaches. These models propose two
different estimations of the reconnection rate~$\varepsilon$: The
Sweet--Parker model predicts $\varepsilon \sim 1/\sqrt{Re_m}$, and
the Petschek model gives $\varepsilon \sim 1/\mbox{ln}{Re_m}$,
where
\begin{equation}
 Re_m = \frac {4\pi V_A L} {c^2\eta}  \label{rem}
\end{equation}
is the global magnetic Reynolds number based on  the half--length of a current
layer $L$, the Alfv\'en velocity $V_A$, and the resistivity of the plasma
$\eta$.
For cosmic plasmas,  magnetic Reynolds numbers usually are very large,
therefore  the
Petschek regime seems to be much more effective. However, since the Petschek
reconnection model was proposed, it is not clear what conditions are necessary
to realize this regime.

It is a fact  that numerical simulations (Biskamp\cite{Biskamp};
Scholer\cite{Scholer}) carried out for a constant resistivity were
not able to reproduce the solution of Petschek type, instead, they
were rather in favour of the Sweet--Parker solution. Laboratory
experiments also seem to observe the Sweet--Parker regime of
reconnection (Ji et al.\cite{Ji}).

On the other hand, if nonuniform resistivity is localized to a
small region, the results of numerical simulations
(Scholer\cite{Scholer}; Ugai\cite{Ugai}) clearly show
Petschek--type reconnection with pronounced  slow shocks. For the
Petschek regime, there are two physically different regions: A
small diffusion region, where dissipation is important, is
surrounded by a large convective zone where the plasma can be
considered as ideal and dissipationless. The problem is very
complicated and thus it does not seem realistic to obtain an
analytical solution which is valid for both regions
simultaneously. To simplify this problem, we seek solutions
separately, in the diffusion region and in the convective zone.
For the later, a solution can be obtained analytically as an
asymptotic series with respect to a small reconnection rate. For
the diffusion region, it is impossible to find an analytical
solution, and hence it has to be obtained numerically. In this
semi--analytical approach, we have to combine the numerical
solution for the diffusion region and the Petschek analytical
solution for the convective region. The latter can be done by
different methods,  which lead to absolutely identical results for
the reconnection rate estimation. The estimation obtained by
Erkaev et al.\cite{Erkaev} is based on asymptotic matching of the
diffusion region and convective zone solutions. In our present
work, we use another way based on a regularized convective region
solution, which seems to be rather clear and very close to the
original Petschek method. In this paper we give a detailed
description of the numerical solution for the diffusion region,
and derive the estimation for the reconnection rate.

This paper is organized as follows: In Sections II and III, we
start with the steady-state MHD equations and present the Petschek
solution. The diffusion region scaling and boundary layer
equations are introduced in Section IV. The numerical algorithm
and the results of the calculations are described in Sections V
and VI. The reconnection rate is derived in Section VII, whereas
Section VIII is devoted to the summary and discussion.
Mathematical details are described in  the Appendix.

\section{MHD equations}

In the problem under consideration, the plasma is governed by the resistive
steady--state MHD system of equations
\begin{eqnarray}
& &\rho(\bf {v \cdot \nabla)\bf v} = -{\bf \nabla} P +
\frac{1}{4\pi}({\bf B \cdot \nabla)\bf B }, \label{1mhd}
\\
& &{\bf E} + \frac{1}{c}({\bf v}\times {\bf B}) =
\frac{c}{4\pi}\eta(x,y)\mbox{curl}\bf {B}, \label{2mhd}
\\
& &\nabla\cdot{\bf B}=0, \quad \nabla\cdot{\bf v}=0, \label{3mhd}
\end{eqnarray}
where $\rho$ is a mass density, $P$ is the total pressure, $P = p
+ B^2/8\pi$, and $Re_m$ is the global magnetic Reynolds number
based on the maximal value of the resistivity $\eta_{max}$.

Outside of the diffusion region, in the so--called convection zone, dissipation
is not important any longer, and we can use the ideal system of MHD equations
in the limit $Re_m \rightarrow \infty$.

In an incompressible plasma the following relations have to be satisfied at the
shock front
\begin{eqnarray}
\left\{ B_n\right\} & = & 0, \label{4mhd}
\\
\left\{ v_n\right\} & = & 0, \label{5mhd}
\\
\left\{ P\right\} & = & 0, \label{6mhd}
\\
\left\{ \frac{1}{4\pi}B_n {\bf B_t}-\rho v_n {\bf v_t} \right\} &
= & 0, \label{7mhd}
\\
\left\{ B_n {\bf v_t}-v_n {\bf B_t} \right\} & = & 0, \label{8mhd}
\end{eqnarray}
where the subscripts $n$ and $t$ denote  components normal and tangential to
the shock front.

\section{\bf Petschek solution}

The Petschek solution, which is valid in the convection region,
can be presented as follows (Petschek\cite{Petschek}, for details
see Vasyliunas\cite{Vasyliunas}). We use coordinates $x,y$, which
are directed along the current sheet and in the perpendicular
direction, respectively. The solution is completely determined by
the following parameters: Quantity $L$ which  is the halflength of
the current sheet, $v_0$ is the plasma inflow velocity, and $B_0$
is the initial magnetic field. The solution is presented in the
form of asymptotic series with respect to the small parameter
which is known as the reconnection rate
\begin{equation}
\varepsilon=\frac{v_0}{V_A} = \frac{E_0}{E_A} \ll 1. \label{Rec}
\end{equation}
Here $E_0$ is the electric field which is constant in the 2D case
under consideration, and $E_A=\frac{1}{c}V_A B_0$ is the Alfv\'en
electric field.

Inflow region:
\begin{eqnarray}
v_x=0\, , &\quad& v_y=-\varepsilon V_A,  \label{Piv}\\
B_x=B_0-\frac{4\varepsilon B_0}\pi \ln \frac L{\sqrt{x^2+y^2}}\, ,&\quad&
 B_y=\frac{ 4\varepsilon B_0}\pi \arctan \frac xy\, . \label{Pib}
\end{eqnarray}

\noindent Outflow region:
\begin{eqnarray}
\phantom{mmmmmm} v_x= V_A, &\quad&  v_y=0\, ,\qquad\\
B_x=0\, ,&\quad& B_y=\varepsilon B_0. \label{Pfr}
\end{eqnarray}
The equation for the shock in the first quadrant is
\begin{equation}
y=\varepsilon x\, . \label{Psk}
\end{equation}
It can be shown that slightly outside of the shock from the inflow side
\begin{equation}\label{Pby}
B_y(x,0)=\left\{  \begin{array}{lll}
 2\varepsilon B_0 x & > & 0\\
 -2\varepsilon B_0 x & < & 0
\end{array} \right.
\end{equation}

Expressions (\ref{Piv}-\ref{Pby}) are  asymptotic solutions with respect to
$\varepsilon$ (zero and first order terms in the inflow region and only zero
order term in the outflow region) of the ideal MHD system of equations
(\ref{1mhd}-\ref{3mhd}) and the Rankine--Hugoniot shock relations
(\ref{4mhd}-\ref{8mhd}).

Petschek did not obtain a solution in the diffusion region, instead, he
estimated the maximum reconnection rate as $1/\mbox{ln}Re_m$ using some simple
physical suggestions. Generally speaking, this implies that the Petschek model
gives any reconnection rate from the Sweet--Parker value $1/\sqrt{Re_m}$ up to
$1/\mbox{ln}Re_m$, and for a long time, it was unclear whether Petschek
reconnection faster than Sweet--Parker reconnection is possible. This problem
can be solved by combining the analytical Petschek solution
(\ref{Piv}-\ref{Pby}) and the numerical model of the diffusion region.

\section{\bf Diffusion region scaling}

The next step is to find a numerical solution for the diffusion
region. But first we have to obtain the boundary layer MHD equations
suitable for the diffusion region.

To this end we renormalize the MHD equations to new scales $B_d,\
V_{Ad},\ E_{Ad}=B_d V_{Ad}/c,\ P_d$, where all quantities are
supposed to be taken at the upper boundary of the diffusion
region,
\begin{eqnarray}
& & x'=x/l_\eta,\quad y'=y/l_\eta,\quad {\bf B}'={\bf B}/B_d,\quad
{\bf v}'={\bf v}/V_{Ad},\quad P'=P/P_d, \label{ld}
\end{eqnarray}
where $l_\eta$ is the characteristic length of the resistivity
variation. The diffusion region length scale $l_d$ (see Figure 1)
obtained from our numerical results (Section VII) is of order of
the scale $l_\eta$.

The convective electric field $-{\bf v\times B}/c\ $ is zero in the centre of
the diffusion region $x=y=0$ where ${\bf v }={\bf B}=0$,  and then increases to
the constant value $E_0$ at the boundary of the convection zone. This type of
behaviour of the convective electric field is reasonable to be used for the
definition of the size of the diffusion region which is one of the most
important parameters of the problem. Namely, the length scale of the diffusion
region is determined as the distance between the origin $x=0$, $y=0$ and the
boundary where the convective electric field reaches its asymptotic value
$E_0$, or better to say, some level, for example, $0.9 E_0$.

In the diffusion region where dissipation is essential, we adopt the
dissipative MHD equations with the magnetic Reynolds number
\begin{equation}
 Re_d' = \frac {4\pi V_{Ad} l_\eta} {c^2 \eta_{max}}\, , \label{redprime}
\end{equation}
and the normalized electric field $E'= E c/ (V_{Ad}B_d)=
\varepsilon'$, where $\varepsilon'$ is a local reconnection rate
at the diffusion region boundary. These electric field and local
reconnection rate are not known. They are to be obtained from the
numerical solution for the diffusion region.

The scaling for the diffusion region is similar to that for the
Prandtl viscous layer (Landau and Lifschitz \cite{Landau}) and
corresponds exactly to the Sweet--Parker one
\begin{equation}
\begin{array}{ll}
& x',\, B'_x,\, v'_x,\, P' \sim O(1),\\
& y',\, B'_y,\, v'_y, \, \varepsilon' \sim 1/\sqrt{Re_d'}\, .
\end{array}
\label{11}
\end{equation}
Consequently, the new boundary layer variables are as follows
\begin{equation}
\begin{array}{ll}
& \tilde{x}=x',\quad \tilde{B}_x=B_x', \quad
\tilde{v}_x=v'_x,\quad \tilde{P}=P',\\
& \tilde{y}=y'\sqrt{Re_d'},\quad
\tilde{B}_y=B_y'\sqrt{Re_d'},\quad
\tilde{v}_y=v_y'\sqrt{Re_d'},\quad
\tilde{\varepsilon}=\varepsilon'\sqrt{Re_d'}.\\
\end{array}\label{12}
\end{equation}

The diffusion region Reynolds number $Re_d'$ is certainly smaller
then the global Reynolds number $Re_m$, but still it is supposed
to be $Re_d'\gg1$. Therefore, in zero--order with respect to  $1/
{Re_d'}$, the boundary layer equations turn out to be
\begin{eqnarray}
& &\frac{\partial \tilde v_x}{\partial t}+ \tilde v{_x}
\frac{\partial \tilde v{_x}}{\partial \tilde x}+
 \tilde v{_y} \frac{\partial \tilde v{_x}}{\partial \tilde y}-
 \tilde B{_x} \frac{\partial \tilde B{_x}}{\partial \tilde x}-
 \tilde B{_y} \frac{\partial \tilde B{_x}}{\partial \tilde y}=
-\frac{\partial \tilde P}{\partial \tilde x}\, , \label{16}
\\
& &\frac{\partial \tilde P}{\partial \tilde y}=0,\label{16a}
\\
& &\frac{\partial \tilde B_x}{\partial t} = \frac{\partial}
{\partial\tilde y} \left(\tilde v{_x}\tilde B{_y}-\tilde
v{_y}\tilde B{_x}\right)+
 \frac{\partial }{\partial \tilde y}\left( \tilde\eta(\tilde x,\tilde y)\frac{\partial
 \tilde B{_x}}{\partial \tilde y}\right)
- \mu \frac{\partial }{\partial \tilde y}\left(\tilde\eta(\tilde
x,\tilde y)\frac{\partial \tilde B_y }{\partial \tilde x} \right)
 ,\label{16b}
\\
& &\frac{\partial \tilde B_y}{\partial t} =
-\frac{\partial}{\partial \tilde x}\left(\tilde v{_x}\tilde
B{_y}-\tilde v{_y}\tilde B{_x}\right)-
 \frac{\partial }{\partial \tilde x}\left( \tilde\eta(\tilde x,\tilde y)\frac{\partial
 \tilde B{_x}}{\partial \tilde y}\right)
 + \mu \frac{\partial }{\partial \tilde
x}\left(\tilde\eta(\tilde x,\tilde y)\frac{\partial \tilde B_y
}{\partial \tilde x} \right)
 ,\label{16c}
 \\
& &\frac{\partial\tilde B_x}{\partial\tilde
x}+\frac{\partial\tilde B_y} {\partial\tilde y}=0\label{17},
\\
& &\frac{\partial\tilde v_x}{\partial\tilde x}+
\frac{\partial\tilde v_y} {\partial\tilde y}=0\label{18},
\end{eqnarray}
where $\tilde \eta(\tilde x,\tilde y)$ is the normalized
resistivity of the plasma with the maximum value to be 1, $\mu$ is
a small parameter, $\mu = 1/Re'_d$. The small terms which include $~\mu$
at the right sides of the induction equations are
necessary for numerical stability of the calculations.

It can be seen from equation (\ref{16a}) that the total pressure
is constant across the diffusion region. This is a general feature
of a boundary layer approximation. Hence, the total pressure is
defined inside the diffusion region by values at the boundary, and
for the boundary layer equations (\ref{16}--\ref{18}), the total
pressure can be considered to be a given function of $x$, e.g.,
$\tilde P(\tilde x)$.

As it was pointed out, the appropriate exact solutions of the boundary layer
equations (\ref{16}--\ref{18}) are unknown even in the steady--state case,
therefore, we have to solve the problem numerically. Although we have to obtain
a steady--state solution, from the point of view of simulation, it is
advantageous to use a relaxation method and to solve numerically the unsteady
system of the boundary layer MHD equations (\ref{16}--\ref{18}).

It is important to note that in the subset of equations
(\ref{16b}--\ref{17}), only two equations are independent. In
principle, we can determine the normal component from the
induction equation (\ref{16c}) or from the equation (\ref{17})
providing the magnetic flux conservation. From the mathematical
point of view, they are equivalent. In our numerical solution, we
use equation (\ref{17}) to determine the $\tilde B_y$ component in
the internal grid points, and the equation (\ref{16c}) is used as
a boundary condition at the lower boundary.

\section{Numerical algorithm}
Starting with an initial MHD configuration under fixed boundary conditions, we
look for the convergence of the time--dependent solution to a steady state. To
avoid additional numerical diffusion, we do not use a flux function and a
magnetic potential. The normalized total pressure is chosen to be 1.

The distribution of the resistivity $\eta =
\eta_{max}\tilde\eta(x,y)$ is traditional (Scholer\cite{Scholer};
Ugai\cite{Ugai})
\begin{equation}
\tilde{\eta}(\tilde{x},\tilde{y})=d e^{(-s_x \tilde{x}^2-s_y
\tilde{y}^2)}+f, \label{eta}
\end{equation}
with $d+f=1$. Setting $d=.95$ and $f=.05$ we can model a case of
localized resistivity, for $d=0$ and $f=1$ the resistivity is
uniform.

As the initial configuration, we choose a current sheet with a
linear profile of the magnetic field $\tilde B_x=\tilde y$,
$\tilde B_y=0$. The velocity components are assumed to be equal to
zero at the initial moment, $\tilde V_x=0$, $\tilde V_y=0$.

To solve the MHD system numerically, we use a two step
conservative finite difference numerical scheme with a rectangular
grid $145\times100$ in the first quadrant. From a time level
($n$), we calculate the parameters on the next time level $(n+1)$
in two steps. In the first step $(n+1/2)$, diffusion is switched
off, and  we calculate the  parameter at the intermediate points
$(n+1/2)$ using the equations in  characteristic form. This is
similar to the approach used in the Godunov method. In the second
step, we calculate the parameters at the next time level $(n+1)$
using the equations in conservative form and taking into account
the diffusion terms approximated in implicit form.

The details of the numerical algorithm  are the following. The
$\tilde B_x$ component is found from the $x$--component of the
induction equation
\begin{eqnarray}
[(B_x)_{i,k}^{n+1}-(B_x)_{i,k}^{n}]/\tau +
(G_{i,k+1/2}^{n+1/2}-G_{i,k-1/2}^{n+1/2})/hx &=&
 \left[\frac{\partial }{\partial \tilde y}\left( \eta(\tilde x,\tilde y)\frac{\partial
 \tilde B{_x}}{\partial \tilde y}\right)\right ]^{n+1}_{i,k}
  \nonumber \\
& &- \mu \left[\frac{\partial }{\partial \tilde
y}\left(\eta(\tilde x,\tilde y)\frac{\partial \tilde B_y
}{\partial \tilde x}
\right)\right]^{n+1}_{i,k},  \label{nu1}
\end{eqnarray}
where the parameters
\begin{eqnarray}
G_{i,k+1/2}^{n+1/2}= (\tilde B_x \tilde V_y- \tilde V_x \tilde
B_y)_{i,k+1/2}^{n+1/2} \label{nu2}
 \end{eqnarray}
are determined by the method of characteristics on the level
$n+1/2$.
This implies that at the beginning ($n \rightarrow
n+1/2$) diffusion is switched  off, and only convection acts, and
then for given convection,
 diffusion is switched on, and $\tilde B_x$ is calculated on the level $n$.
The normal magnetic field component $\tilde B_y$ is determined from the
equation $\nabla\cdot{\bf B} =0$

The velocity component $\tilde V_x$ is found from the
$x$--component of the momentum equation (\ref{16}),
\begin{eqnarray}
[(\tilde V_x)_{i,k}^{n+1}-(\tilde V_x)_{i,k}^{n}]/\tau + (Q_{y i,
k+1/2} - Q_{y i, k-1/2})^{n+1/2}/hy+ (Q_{x i+1/2, k} - Q_{y i-1/2,
k})^{n+1/2}/hx = 0, \label{nu4}
\end{eqnarray}
 where
\begin{eqnarray}
& Q_{y i, k+1/2}^{n+1/2}=(\tilde V_x \tilde V_y - \tilde B_x
\tilde B_y )_{i,k+1/2}^{n+1/2}\, ,\label{nu5}
\\
&Q_{x i+1/2, k}^{n+1/2}=(V_x^2 -
B_x^2)_{i+1/2,k}^{n+1/2}\, .\label{nu6}
\end{eqnarray}
Here, the parameters $( )_{i,k+1/2}^{n+1/2}\, $ are determined by
the method of characteristics on the level $n+1/2$ simultaneously
with the calculation of $\tilde B_x$. The velocity component
$\tilde V_y$ is determined from the equation $ \mbox{div}{\bf
V}=0$.

The boundary conditions are as follows: \\
At the upper (inflow)
 boundary,
the tangential magnetic field component is assumed to be constant,
$\tilde B_x$ =1 and the tangential velocity component vanishes $\tilde V_x =0$. \\
At the left boundary we have the symmetry conditions,
$\partial \tilde B_x /\partial \tilde x =0$, $\tilde B_y =0$, $\tilde V_x =0$. \\
At the right boundary we hold  free conditions suitable for a
uniform flow in the outflow region, $\partial \tilde B_y /\partial
\tilde
x =0$, $\partial \tilde V_y /\partial \tilde x =0$.

At the lower boundary ($y=0$) there is the symmetry condition for
the tangential magnetic field component, $\tilde B_x = 0$, and the
non--flow condition for the normal velocity component, $\tilde V_y
= 0$. At this boundary, the normal component of the magnetic field
$\tilde B_y$ is obtained from the induction equation (\ref{16c})
on the line $y=0$,
\begin{eqnarray}
  \frac{\partial \tilde B_y}{\partial t} + \frac{\partial}{\partial
t}(\tilde V_x \tilde B_y)=-
 \frac{\partial }{\partial \tilde x}\left( \eta(\tilde x,\tilde y)\frac{\partial
 \tilde B{_x}}{\partial \tilde y}\right)
 + \mu \frac{\partial }{\partial \tilde
x}\left(\eta(\tilde x,\tilde y)\frac{\partial \tilde B_y
}{\partial \tilde x} \right) .\label{nu3}
\end{eqnarray}
The small parameter $\mu \sim .1-.2$ is used here to regularize
the numerical scheme for the  unsteady system of the boundary
layer MHD equations (\ref{16}--\ref{18}), which is an ill--posed
problem in our case.

The size of the computational domain is chosen to be much larger
than the diffusion region size $l_d$, and also much less than the
global size $L$. At the inflow boundary we do not fix the normal
components of the magnetic field and velocity, and thus we do not
impose a reconnection rate and an electric field in the diffusion
region from the very beginning. The latter has to be found from
the numerical solution self--consistently.

\section{Results of the numerical simulation}

To estimate the convergence of the time--dependent solution to a steady
state for each $n$--th time step, we use the following criteria,
$\mbox{max}(|V_{x}^{n}-V_{x}^{n-1}|)/({\Delta t} |V_{x max}^{n}|)< 10^{-6}$.
In the 2D steady state the total (convective plus dissipative) electric
field must be constant, and it is so in our simulations (see
Figures 2, and 3) besides of small perturbations near the outflow
boundary due to some reflections, although we apply free boundary
conditions.

Let us discuss the result of our simulations. For the case of
localized resistivity, the system reaches the Petschek steady
state with clear asymptotic behaviour (see Figure 2): $\tilde V_x
\rightarrow 1$ in the outflow region; $\tilde V_y \rightarrow
\tilde\varepsilon$ at the inflow boundary; $\tilde B_x$ decreases
from 1 to 0 at the shock transition; $\tilde B_y \rightarrow
\tilde\varepsilon$ in the outflow region; and $\tilde B_y
\rightarrow 2 \tilde \varepsilon$ from the inflow side of the
shock (compare with the Petschek solution (\ref{Piv}--\ref{Psk})).

There is a well pronounced slow shock, as can be seen in the behaviour of all
MHD parameters, but in particular in the distribution of the current density.
The normalized electric field (reconnection rate) turns out to be
$\tilde\varepsilon \sim 0.7$. It is important to note that the numerical
results do not depend on the size of calculation box.

On the other hand, for the case of homogeneous resistivity, the
system reaches the Sweet--Parker state (see Figure 3), even if the
Petschek solution is used as initial configuration (see also
Scholer\cite{Scholer}; Ugai\cite{Ugai}; Uzdensky and
Kulsrud\cite{Uzdensky}). This seems to imply that Petschek--type
reconnection is possible only if the resistivity of the plasma is
localized to a small region, whereas for constant resistivity, the
Sweet--Parker regime is realized (Erkaev et al.\cite{Erkaev}).

The size of the diffusion region layer $l_d$ is defined as its
length along the $x$ axis where the convective electric field at
the lower boundary ($y = 0$) ${\tilde E_c}=-{\tilde v_x} {\tilde
B_y}$ is less in absolute value than some level of the total
electric field (say $0.9 \tilde\varepsilon$). For the case of a
localized resistivity, $l_d$ practically coincides with the scale
of the inhomogeneity of the resistivity $l_\eta$ when the maximum
of resistivity is much larger then the background resistivity.
Therefore hereafter we consider $l_d \sim l_\eta$.

For the case of uniform resistivity, the plasma is accelerated
very slowly, and there is no obvious definition for the scale
length of the diffusion region. Diffusion is important everywhere
for the pure Sweet--Parker regime, and for the Petschek asymptotic
solution there is left no room. Therefore, the solution does not
converge to the Petschek solution,  not only at the right hand
boundary but everywhere. In this case, the solution will depend
on the calculation box size because it does not have any other
scale. Hence, the constant resistivity solution can not be matched
to the Petschek solution.

Nevertheless, the Sweet--Parker regime is still important also for the Petschek
solution, because in the nearest vicinity of the reconnection line, where the
resistivity can be considered to be constant, the diffusion region structure is
similar to the Sweet--Parker case. Besides, and this is even more important,
the scaling for the diffusion region is exactly the Sweet--Parker one
(\ref{11}, \ref{12}), or, better to say, the Prandtl scaling.

\section{Reconnection rate}

To find a relationship between the reconnection rate and
dissipation we need first of all an estimation of magnetic field
at the boundary of the diffusion region $B_d$. To this end we can
not use the Petschek solution (\ref{Pib}) because the $B_x$
component diverges at the origin $B_x\rightarrow -\infty $, when
${r=\sqrt{x^2+y^2}}\rightarrow 0$. This singularity is a
consequence of the fact that dissipation actually has not been
taken into account for the Petschek solution. Formally it follows
from the jump at the origin of the $B_y$ component of the magnetic
field (\ref{Pby}). Dissipation evidently leads to smooth behaviour
of the magnetic field in the diffusion region, and then no
singularities are possible. To illustrate this we consider a model
distribution of the $B_y(x,0)$ component with linearly smoothed
boundary condition at the interval $(-l_d,l_d) $ similar to the
original Petschek\cite{Petschek} consideration
\begin{equation}\label{regcon}
 B_y^{P}(x,0)=
 \left\{ \begin{array}{lll}
     \pm 2\varepsilon B_0 &L>|x|& >l_d\\[1mm]
      2\varepsilon B_0 \frac{x}{l_d} &|x|< l_d& \\[1mm]
      0 &|x|>L &\, .
      \end{array}
 \right.
\end{equation}

The $B_x(x,y)$ component of the magnetic field in the inflow
region can be found from the Poisson integral,
\begin{eqnarray}
B_x(x,y) & = &
B_0-\frac{1}{\pi}\int\limits_{-\infty}^{+\infty}\frac{B_y^{P}(x',0)(x'-x)}
     {(x'-x)^2+y^2}dx' \nonumber\\
      & = & B_0-\frac{2\varepsilon B_0}{\pi l_d}\left(2l_d+\frac{x}{2}\ln\frac{(x-l_d)^2+y^2}
      {(x+l_d)^2+y^2}+
      y\arctan\frac{x-l_d}{y}-y\arctan\frac{x+l_d}{y}\right)-\nonumber\\
      & & \vphantom{\int\limits_0^0}-\frac{\varepsilon B_0}{\pi}\left(\ln
      \frac{(y^2+(L-x)^2)(y^2+(L+x)^2)}{(y^2+(l_d-x)^2)(y^2+(l_d+x)^2)}\right).
      \label{bxr}
\end{eqnarray}
This solution does not have a singularity at the origin any more,
and tends to the Petschek solution (\ref{Pib}) outside the
diffusion region. We can simplify equation (\ref{bxr}) at the
origin
\begin{equation}
   B_x(0,0)=
      B_0-\frac{4\varepsilon B_0}{\pi}\ln\frac{L}{l_d}-\frac{4\varepsilon B_0}
      {\pi}.
      \label{bx00}
\end{equation}
The first term on the right hand side of this equation is
of the oder of $O(1)$, the third one is of  $O(\varepsilon)$, but the
second term consists of a large parameter $\ln\frac{L}{l_d}$ times
the small parameter $\varepsilon$. Thus we assume the following
relations between the parameters
\begin{equation}
  1 > \varepsilon\ln\frac{L}{l_d}\gg\varepsilon. \label{parier}
\end{equation}
So far we considered only a model distribution of the $B_y^P(x,0)$
(\ref{regcon}) along the current sheet but it turns out that
$B_x(0,0)$ does not depend on the actual distribution of the $B_y$
component inside the diffusion region up to $O(\varepsilon)$. This
implies that we can extend equation (\ref{bx00}) to the general
case.

Let us consider the Poisson integral with the actual distribution of the
$B_y(x,0)$
component using the model boundary condition $ B_y^{P}(x,0)$ (\ref{regcon}) for
regularization
\begin{eqnarray}
B_x(0,0) & = &
B_0-\frac{1}{\pi}\int\limits_{-\infty}^{+\infty}\frac{B_y(x',0)}
      {x'}dx' \nonumber\\
      & = & B_0-\frac{1}{\pi}\int\limits_{-\infty}^{+\infty}\frac{\left(B_y(x',0)-
      B_y^{P}(x',0)+B_y^{P}(x',0)\right)}{x'}dx'\nonumber\\
      & = & B_0-\frac{4\varepsilon B_0}{\pi}\ln\frac{L}{l_d}-\frac{4\varepsilon B_0}
      {\pi}+
      \frac{1}{\pi}\int\limits_{-\infty}^{+\infty}\frac{\left(B_y(x',0)-
      B_y^{P}(x',0)\right)}{x'}dx'\nonumber\\
      & = & B_0-\frac{4\varepsilon B_0}{\pi}\ln\frac{L}{l_d}- C\varepsilon B_0,
      \label{bxreg}
\end{eqnarray}
where $C=\mbox{const}\ $  includes both, the contribution from
$\frac{4\varepsilon B_0}{\pi}$ and the contribution from the
non--singular integral in the third line of this equation. The
main difficulty for the estimation of this integral is that near
the diffusion region, the local Petschek solution reproduced in
our simulation, seems to be different from the global one because
$\varepsilon'>\varepsilon$ and $B_d<B_0$. The local Petschek
solution has asymptotically $B_y(x/l_d)\rightarrow 2\varepsilon'
B_d$ when $x/l_d\rightarrow \infty$ which seems to be different
from the condition $B_y^P(x/l_d)\rightarrow 2\varepsilon B_0$ used
in (\ref{regcon}). However, as it is shown in Appendix, the
difference $O(\varepsilon')-O(\varepsilon)$ is of the order of
$\varepsilon$ rather then $O(\varepsilon\ln\frac{L}{l_d})$ (see
Appendix). This allows us to estimate the integral (\ref{bxreg})
as a quantity of order $\epsilon$ which is much smaller than the
main term $\sim \varepsilon\ln\frac{L}{l_d}$.

The diffusion region is small $l_d\ll L$ and for the boundary condition for the
diffusion region $B_d$ we can use the magnetic field at the origin $B_x(0,0)$.
Using the relation (\ref{bxreg}), we find the magnetic field strength at the
diffusion region boundary
\begin{eqnarray}
B_d=B_0(1-\frac{4\varepsilon}\pi \ln \frac L{l_d})\, . \label{22}
\end{eqnarray}
 Now everything is ready to determine the reconnection rate. The
electric field must be constant in the whole inflow region, hence
\begin{eqnarray}
&v_d B_d=v_0B_0, \label{23}
\\
&\varepsilon' B_d^2 = \varepsilon B_0^2 , \label{24}
\end{eqnarray}
where the definition of the reconnection rates $\varepsilon'=v_d/B_d,\ \
\varepsilon=v_0/B_0$ are used. Bearing in mind that $\varepsilon
'=\tilde\varepsilon/\sqrt Re_d'$ (see scaling (\ref{12})) we obtain
\begin{equation}
\tilde\varepsilon{B_d}^{3/2}= \varepsilon B_0^{3/2} \sqrt{\frac{4 \pi
V_{Ad}l_d}{c^2\eta_{max}}} \, . \label{25}
\end{equation}
Substituting $B_d$ from equation (\ref{22}), we determine finally the following
equation for the reconnection rate $\varepsilon$
\begin{equation}
\tilde\varepsilon(1-\frac{4\varepsilon}\pi \ln \frac
L{l_d})^{3/2}= \varepsilon \sqrt{Re_d}\, , \label{26}
\end{equation}
where the magnetic Reynolds number $Re_d=4\pi V_A
l_d/(c^2\eta_{max}) $ is based on the global Alfv\'en velocity and
the half length of the diffusion region $l_d$. The internal
reconnection rate $\tilde\varepsilon$ has to be found from the
simulation of the diffusion region problem.

For small $\varepsilon\ln\frac{L}{l_d}$ there is an analytical
expression
\begin{equation}
\varepsilon=\frac{\tilde\varepsilon}{\sqrt{Re_d}+ \frac 6\pi
\tilde\varepsilon\ln \frac L{l_d}}\: . \label{27}
\end{equation}
Here $ \tilde\varepsilon$ is an internal reconnection rate,
determined from the numerical solution, which is $\tilde\varepsilon \sim
0.7$ for the Petschek type solution.

In the Appendix it is also shown that the global Petschek solution
with  second order corrections tends to the asymptotic of the
diffusion region solution for $x \sim l_d$.

It is interesting that for the derivation of the final result (\ref{26},
\ref{27}) the only value which has been actually used is the internal
reconnection rate $\tilde\varepsilon$ obtained from the numerical solution, and
the asymptotic behaviour (\ref{regcon}). The actual distribution of the $B_y$
component along the upper boundary of the diffusion region does not contribute
at all (besides of the asymptotic behaviour (\ref{regcon})) in  zero--order
approximation considered above. Of course, from the mathematical point of view,
it is important that the diffusion region solution exists and has the
Petschek--like asymptotic behaviour (\ref{Piv}--\ref{Pby}). Therefore, the
asymptotic behaviour (\ref{regcon}) plays the key role in the derivation of the
reconnection rate  and this question needs to be clarified in more detail.

\section{\bf Discussion }

Equations (\ref{26}, \ref{27}) give the unique reconnection rate
for known parameters of the current sheet  $L,$ $ B_0,$ $ V_A,$ $
\eta,$ $ l_d$. Let us fix now the lengths $L$ and start to vary
$l_d$ assuming $l_\eta\sim l_d$. It is clear that for small $l_d$,
the Petschek term becomes large, whereas for big $l_d$, the
Sweet--Parker term is dominant.
The behavior of the implicit
function $\varepsilon(l_d/L)$ given by (\ref{26}) is non
monotonic. There exists a length $l_d$ corresponding to a maximum
value of the reconnection rate. This maximal reconnection rate is
a function of the magnetic Reynolds number given in an implicit
form
\begin{equation}
   \varepsilon = \frac{\pi}{4 (A + \ln(Re_m/\varepsilon))},
\end{equation}
where $A$ is the constant $A = 3 - 2 \ln(\tilde\varepsilon) -3 \ln(12/\pi) =
-0.31$. Here $Re_m$ is the Reynolds number determined for the global scale and
the maximal resistivity $Re_m = 4\pi V_A L/(c^2 \eta_{max})  $. This result can
be interpreted as follows. In the case of a large global Reynolds number, for
fixed values of the maximum resistivity and the global scale $L$, the
reconnection rate and the corresponding intensity of energy conversion reach
their maxima when the diffusion region length scale and also the conductivity
length scale are much smaller than $L$. This maximum value of the reconnection
rate is a logarithmic function of the global Reynolds number which is similar
to that estimated by Petschek. This fact contradicts to the usual
electrotechnical intuition. For example, to get maximum heating from a rheostat
(resistor), we need to switch on the whole length, to increase $l_d$, as oppose
to the progress of reconnection. It is a fact that the energy release in the
course of the reconnection process takes place not only in the form of Joule
heating in the diffusion region and at the shock fronts, but also in the form
of plasma acceleration.

By increasing the conductivity length scale and
the corresponding diffusion region length scale,
the reconnection rate decreases substantially,
becoming more close to that of the Sweet--Parker regime.

We have to emphasize once more that the case of constant
resistivity is not described by equation (\ref{27}), because there
is no clear scale of the diffusion region, no clear Petschek--type
asymptotic behaviour, and therefore it can not be matched with the
Petschek solution.

The appearance of strongly localized resistivity is often the
relevant case in space plasma applications, but for  laboratory
experiments, where the size of a device is relatively small, the
Petschek regime can hardly be expected.

One of the main difficulties of the diffusive--like theories of
reconnection such as the Sweet--Parker mechanism (Sweet, 1958,
Parker, 1963), and the tearing instability (Galeev et al., 1986)
is that the efficiency of the process turns out to be of the order
of $Re_m^{-\alpha}$ where usually $0<\alpha<1$. For example, for
the Sweet--Parker regime, $\alpha=1/2$. In cosmic plasmas the
magnetic Reynolds number is often very large because of the large
scale, high velocity and high conductivity. Hence, the efficiency
of pure dissipative processes is rather poor. The Petschek
mechanism of fast reconnection is much more effective due to the
logarithmic dependence of the reconnection rate on scale
(42). In the Petschek model, MHD waves play the dominant role
and the logarithmic dependence is the contribution of the waves to
the efficiency of the process.

In this paper, we studied reconnection for a strongly localized resistivity
with a large ratio of the maximal and background resistivity (~20). A crucial
parameter for the reconnection rate is the diffusion region length which is
obtained to be approximately equal to the length scale of the resistivity. An
interesting question for future study is the dependence of the diffusion region
length as well as the electric field on the amplitude of the resistivity
variation.

\section{\bf Appendix}

So, we have to clarify the problem concerning the asymptotic behaviour
$B_y(x/l_d)\rightarrow 2\varepsilon' B_d$ when $x/l_d\rightarrow \infty$,
estimate the integral, and to prove that the global Petschek solution tends to
the local one if we take into account all necessary terms. Originally, Petschek
(1964) considered the reconnection problem using as a small parameter the
reconnection rate $\varepsilon$. He obtained the solution
(\ref{Piv}-\ref{Pby}), taking into account only zero and first order terms in
the inflow region, and zero order terms in the outflow region. But there is the
possibility to extend this solution with higher order terms (Pudovkin and
Semenov, 1985). In order to do this we have to present each component of the
MHD state vector $U$ (inflow region), $\hat U$ (outflow region), $S$ (shock
front ) as an asymptotic series with respect to the reconnection rate
$\varepsilon$
\begin{eqnarray}
U&=&U^{(0)}+\varepsilon U^{(1)}+\varepsilon^2 U^{(2)}+\ldots\label{absser_i}
\\
\hat U&=&\hat U^{(0)}+\varepsilon\hat U^{(1)}+\varepsilon^2\hat
U^{(2)}+\ldots\label{absser_o}
\\
S&=&S^{(0)}+\varepsilon S^{(1)}+\varepsilon^2 S^{(2)}+\ldots\label{absser_s}
\end{eqnarray}
 The terms of the series
(\ref{absser_i}-\ref{absser_s}) can be obtained step by step using the MHD
equations (\ref{1mhd}-\ref{3mhd}) and the shock boundary conditions
(\ref{4mhd}-\ref{8mhd}) according to the following scheme,
\begin{equation}
U^{(0)}\mathop{\Rightarrow}\limits^{\it1} \hat U^{(0)}
       \mathop{\Rightarrow}\limits^{\it2} S^{(0)}
       \mathop{\Rightarrow}\limits^{\it3} U^{(1)}
       \mathop{\Rightarrow}\limits^{\it4} \hat U^{(1)}
       \mathop{\Rightarrow}\limits^{\it5} S^{(1)}
       \mathop{\Rightarrow}\limits^{\it6}\ldots
\end{equation}
Here $U^{(0)}$ is the initial vector, and each next term is determined  via
solving the reduced MHD system with boundary condition provided by the previous
step.

For example, the original Petschek solution (\ref{Piv}-\ref{Pby}) corresponds
to the first three steps of this scheme. The first step is trivial, because no
shock front is yet possible. In the next step, the outflow region solution of
zero order allows to impose a boundary condition problem for the inflow region
solution in first order, and so on.

Proceeding according to this scheme up to the step $\it5$, we obtain the
following extended Petschek solution.

Inflow region:
\begin{eqnarray}
B_x&=&B_0-\frac{4\varepsilon}{\pi}B_0\ln\frac{L}{\sqrt{x^2+y^2}}\label{ibx}\label{pfrom}\label{inffrom}\\
B_y&=&\frac{4\varepsilon}{\pi}B_0\arctan\frac{x}{y}\label{iby}\\
V_x&=&\frac{4\varepsilon}{\pi}V_0\arctan\frac{x}{y}\label{ivx}\\
V_y&=&-V_0-\frac{4\varepsilon}{\pi}V_0\ln\frac{L}{\sqrt{x^2+y^2}}\label{ivy}\label{infto}
\end{eqnarray}

Outflow region:
\begin{eqnarray}
 B_x&=&\frac{4\varepsilon}{\pi}B_0\ln\frac{x+\hat y}{x-\hat
y},\label{obx}\label{offrom}
\\
 B_y&=&\varepsilon
  B_0-\frac{4\varepsilon^2}{\pi}B_0\ln\frac{x^2-\hat y^2}{4xL},\label{oby}
  \\
 V_x&=&V_A+\frac{4V_0}{\pi}\ln\frac{x^2-\hat
y^2}{4Lx},\label{ovx}
\\
 V_y&=&\frac{4\varepsilon V_0}{\pi}\left(\ln\frac{x+\hat
y}{x-\hat y}+\frac{\hat y}{x}\right),\label{ovy}\label{pto}\label{ofto}
\end{eqnarray}
where $\hat y=y/\varepsilon$.

Shock front equation:
\begin{equation}
y=\varepsilon
x+\frac{4\varepsilon^2}{\pi}\left(2x\ln\frac{x}{L}+x\right).\label{osk}
\end{equation}

Finally it is possible to find the $y$--component of the magnetic field ${ B_y
(x)}$ at the shock which has been used in deriving the reconnection rate up to
second order,
\begin{equation}
 B_y= 2 B_0 \varepsilon \left(1-\frac{4 \varepsilon
}{\pi}(\ln\frac{x}{L}+3)\right). \label{obyb}
\end{equation}

Using the extended Petschek  solution (\ref{pfrom}--\ref{obyb}), we can prove
now that the global solution tends to the local one at $x \sim l_d$. From
equations (\ref{22}, \ref{24}) it follows that
\begin{equation}
\varepsilon'=\varepsilon B_0^2 \left( 1+\frac{8 \varepsilon }{\pi}
\ln\frac{L}{l_d}\right).\label{eps}
\end{equation}

Let us check now that $B_y(x)\rightarrow 2\varepsilon' B_0'$ for $x \sim l_d$
at the inflow side of the shock. On one hand, we can expect that near the
diffusion region
\begin{equation}
B_y'=2\varepsilon' B_0'= 2 B_0 \varepsilon \left(1+\frac{8 \varepsilon
}{\pi}\ln\frac{L}{l_d}\right)\left(1-\frac{4 \varepsilon
}{\pi}\ln\frac{L}{l_d}\right)=2 B_0 \varepsilon \left(1+\frac{4 \varepsilon
}{\pi}\ln\frac{L}{l_d}\right).\label{1by}
\end{equation}
On the other hand, for $x \sim l_d$, the global solution tends to
\begin{equation}
 B_y= 2 B_0 \varepsilon \left(1-\frac{4 \varepsilon
}{\pi}(\ln\frac{x}{L}+3)\right)_{x=l_d}=2 B_0 \varepsilon \left(1+\frac{4
\varepsilon }{\pi}\ln\frac{L}{l_d}\right).\label{2by}
\end{equation}
Therefore $B_y(x)\rightarrow 2\varepsilon' B_0'$, if we take into account the
next term in the $\varepsilon$ expansion for $B_y$ at the shock. This resolves
the question concerning the asymptotic behaviour $B_y(x/l_d)\rightarrow
2\varepsilon' B_d$ when $x/l_d\rightarrow \infty$.

Similarly it can be shown that the global Petschek solution tends to the local
one at the distance $x \sim l_d$. This implies that all components of $\bf V$,
$\bf B$ are matched automatically near the boundary with the convection zone if
one of them ($B_x$ in our case) has been adjusted properly.

Now we can estimate the integral used in equation (\ref{regcon}):
\begin{equation}
      \frac{1}{\pi} \int\limits_{-L}^{L}\frac{\left(B_y(x',0)-
      B_y^{P}(x',0)\right)}{x'}dx'=\frac{1}{\pi}\int\limits_{-L}^{-l_d}+
      \frac{1}{\pi}\int\limits_{-l_d}^{l_d}+\frac{1}{\pi}\int\limits_{l_d}^{L}
      \label{int}
\end{equation}
The integral over the diffusion region $x\in (-l_d, l_d)$ is estimated as
$O(\varepsilon)$ since   $B_y(x,0)- B_y^{P}(x,0)$ is an odd function of $x$,
and the integral converges in the usual sense rather than to be calculated as a
principal value. The contribution from the intervals $(-L,-l_d)$ and $(l_d,L)$
are estimated as $O(\varepsilon^2 \mbox{ln}(L/l_d))$ because as it follows from
equation (\ref{2by}), the difference $B_y(x,0)- B_y^{P}(x,0) \sim
O(\varepsilon^2 \mbox{ln}(L/l_d))$. Taking into account the hierarchy of the
small
parameters (\ref{parier}) we conclude that the whole integral (\ref{int}) is
estimated as $O(\varepsilon)$.

\section{Acknowledgements}

\hspace*{8mm}This work is supported by the INTAS-ESA project 99-01277. It is
also supported in part by grants No \mbox{01-05-65070} and No
\mbox{01-05-64954} from the Russian Foundation of Basic Research and by the
programme ``Intergeophysics'' from the Russian Ministry of Higher Education.
Part of this work is supported by the ``Fonds zur F\"orderung der
wissenschaftlichen Forschung'', project P13804-TPH. This work is further
supported by grant No \mbox{01--05--02003} from the Russian Foundation of Basic
Research and by project I.4/2001 from ``\"Osterreichischer Akademischer
Austauschdienst''. We acknowledge support by the Austrian Academy of Sciences,
``Verwaltungsstelle f\"ur Auslandsbeziehungen''.

\newpage

{\large {\bf  Figure Captions }}

Figure 1: {Scheme of Petschek reconnection.}

Figure 2: {Numerical results for  Petschek-type reconnection with localized
resistivity. Left column: structure of magnetic field lines (solid lines) and
stream lines (dashed), distributions of the $V_x, B_x$, and convection electric
field. Right column: distributions of the electric current, $V_y, B_y$, and
total electric field.}

Figure 3: {Numerical results for  Sweet--Parker reconnection with constant
resistivity. Left column: structure of magnetic field lines (solid lines) and
stream lines (dashed), distributions of the $V_x, B_x$, and convection electric
field. Right column: distributions of the electric current, $V_y, B_y$, and
total electric field.}

\vfill

\end{document}